\documentclass[twocolumn,prb,showpacs]{revtex4}
\usepackage{graphicx}
\usepackage{dcolumn}
\usepackage{bm}
\usepackage{amsmath}

\begin{document}

\preprint{}
\title{ Spin-spin interaction mediated by chiral phonons}
\author{Takehito Yokoyama}
\affiliation{Department of Physics, Tokyo Institute of Technology, Tokyo 152-8551,
Japan 
}
\date{\today}

\begin{abstract}
We study interaction between two magnetic impurities on top of a two dimensional insulator in the presence of chiral phonons by second-order perturbation theory.
We show that this exchange interaction arises from angular momentum of phonons through spin-chiral phonon interaction of the form of spin-orbit coupling.
Analytical expressions of the interaction for acoustic and optical phonons are obtained. The exchange interactions are always positive due to the bosonic nature of phonons.
We find that the exchange interactions for acoustic and optical phonons show power-law decay with respect to the distance between the two magnetic impurities and are proportional to the temperature of the system at high temperature.

\end{abstract}

\maketitle



Chiral phonons are circularly polarized vibrational motions of ions in solids.\cite{LZhang2014,LZhang2015,Chen2019,Komiyama,Wang} Due to the rotational motion of the ions, chiral phonons exhibit angular momentum and magnetic moment.\cite{HZhu2018,Ishito2023,Grissonnanche2020,XTChen2019,Li2019,Ueda2023}
The rotation of the ions also induces  effective magnetic fields~\cite{Nova2017,Juraschek2019,Geilhufe,Juraschek2022,Xiong2022,Luo2023,Hernandez2022,Chaudhary2023} which have been estimated as 0.01T in tellurium\cite{Xiong2022}, 100 T for cerium trichloride~\cite{Juraschek2022}, and one T in cerium fluoride~\cite{Luo2023}. With these magnetic field, one can expect a coupling between chiral phonons and other degrees of freedom such as spins which has been extensively investigated in recent years.
In fact, couplings between spins and chiral phonons lead to conversion between electron spin and microscopic atomic rotation~\cite{HamadaPRR2020}, conversion of chiral phonons into magnons\cite{Yao2023},
ultrafast demagnetization\cite{Tauchert2022}, magnetization manipulation or reversal\cite{Basini,Kahana2023,Davies2023},
light driven spontaneous phonon chirality and magnetization\cite{YafenRen2023},
interactions between chiral phonons and eletric spin\cite{Fransson2023} or orbital\cite{Ren2021} magnetizations, and spin current generated by chiral phonons\cite{Kim2023,Li2022,Yao,Funato}.

Recent experimental reports have suggested spin-spin interaction mediated by chiral phonons. 
Long-range exchange interaction through  more than 30-nm-thick insulator has been reported in Refs. \cite{Korenev,Korenev2}.
It is conjectured that this long-range exchange coupling is mediated by elliptically polarized phonons.\cite{Korenev,Korenev2}
Chiral phonon-mediated interlayer exchange interaction in ferromagnetic metal-nonmagnetic insulator heterostructures has been also investigated experimentally in Ref.\cite{Jeong2022}. It has been reported that spins rotate as a function of nonmagnetic insulating spacer thickness.\cite{Jeong2022}
However, although it is plausible that this interlayer exchange interaction stems from chiral phonons,\cite{Jeong2022} its microscopic mechanism still remains to be clarified.
As for mechanisms of spin-spin interactions, electronic origins have been mostly investigated.\cite{RKKY,Bruno} Thus, one fundamental question is: what is a signature of  spin-spin interactions mediated by bosons?

To address these problems, in this paper,
we study interaction between two localized spins on top of a two dimensional insulator in the presence of chiral phonons by second-order perturbation theory.
We show that this interaction is mediated by chiral phonons through spin-chiral phonon coupling of the form of spin-orbit coupling.
Analytical expressions of the interaction for acoustic and optical phonons are obtained. The exchange interactions are always positive due to the bosonic nature of phonons.
We find that the exchange interactions for acoustic and optical phonons show power-law decay with respect to the distance between the two localized spins and are proportional to the temperature of the system at high temperature.

\begin{figure}[htb]
\begin{center}
\includegraphics[clip,width=8.0cm]{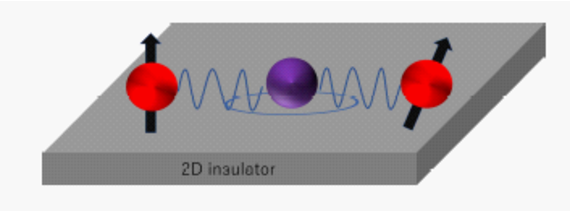}
\end{center}
\caption{
Schematic diagram of the model. Two magnetic impurities on top of a two dimensional insulator interact with each other by chiral phonons.}
\label{f1}
\end{figure}

We consider two localized spins on a two dimensional insulator in the presence of chiral phonons as shown in Fig. \ref{f1}.
The Hamiltonian for phonons $H_0$ reads
\begin{eqnarray}
{H_0} = \sum\limits_{{\bf{q}}\nu} {\omega _{\bf{q}\nu}} a_{\bf{q}}^{\nu\dag }a_{\bf{q}}^{\nu }
\end{eqnarray}
where $a_{\bf{q}}^{\nu \dag } (a_{\bf{q}}^{\nu})$ is the creation (annihilation) operator for a phonon with wave vector $\bf{q}$ and mode $\nu$ with the corresponding phonon dispersion $\omega _{\bf{q}\nu}$.
We set $k_B=\hbar=1$ throughout this paper. 

We define the phonon Green function $D$ by
\begin{eqnarray}
D({\bf{q}},\tau ,\nu) =  - \left\langle {{{\rm{T}}_\tau }\left[ {{\varphi _{{\bf{q}}\nu}}(\tau )\varphi _{{\bf{q}}\nu}^\dag } \right]} \right\rangle 
\end{eqnarray}
with $\varphi _{{\bf{q}}\nu}^\dag  = \frac{1}{{\sqrt 2 }}\left( {a_{\bf{q}}^\nu + a_{ - {\bf{q}}}^{\nu\dag }, - i\left( {a_{\bf{q}}^{\nu } - a_{ - {\bf{q}}}^{\nu}\dag } \right)} \right)$.
These two components correspond to the second quantization representation of the displacement and momentum of atoms.

We perform a Fourier transform of the phonon Green function $D({\bf{q}},\tau ,\nu)$ for $H_0$ and obtain 
\begin{eqnarray}
D({\bf{q}},\omega _n^{},\nu) = \frac{1}{{\omega _n^2 + \omega _{{\bf{q}}\nu}^2}}\left( {\begin{array}{*{20}{c}}
{ - \omega _{{\bf{q}}\nu}^{}}&{\omega _n^{}}\\
{ - \omega _n^{}}&{ - \omega _{{\bf{q}}\nu}^{}}
\end{array}} \right) \nonumber \\
\equiv D_0({\bf{q}},\omega _n^{},\nu) \tau_0 +  D_2({\bf{q}},\omega _n^{},\nu) i \tau_2 
\end{eqnarray}
with the Matsubara frequencies $\omega _n=2\pi n T$ and the Pauli matrices $\tau_0$ and $\tau_2$. Here, $T$ and $n$ are the temperature and an interger, respectively. 

The angular momentum of phonons along $z$ direction, i.e., direction perpendicular to the 2D insulator, can be expressed by the second quantization representation of the displacement $u$ and momentum $p$  as \cite{Gao2023}
\begin{eqnarray}
{L^z} = u^xp^y - u^yp^x = \frac{1}{{2}}\sum\limits_{{\bf{q}},\nu} {g_{\bf{q}}^{\nu ,{\nu^\prime }}\left( {a_{\bf{q}}^\nu + a_{ - {\bf{q}}}^{\nu\dag }} \right)} \left( {a_{\bf{q}}^{\nu '\dag } - a_{ - {\bf{q}}}^{\nu'}} \right)
\end{eqnarray}
with
\begin{eqnarray}
g_{\bf{q}}^{\nu ,{\nu^\prime }} = \sqrt {\frac{{\omega _{\bf{q}\nu^\prime }}}{{\omega _{\bf{q}\nu}}}} \xi _{{\bf{q}},{\nu ^\prime }}^\dag \left( {\begin{array}{*{20}{c}}
0&{ - i}\\
i&0
\end{array}} \right){\xi _{{\bf{q}},\nu }}.
\end{eqnarray}
Here, ${\xi _{{\bf{q}},\nu }}$ is the polarization vector. We assume doubly degenerate acoustic and doubly degenerate optical modes which may be realized in hexagonal lattices.\cite{LZhang2014}  Each of circular polarized phonon modes is labeled by $\nu=\pm$. Now, we assume
${\xi _{{\bf{q}},\nu }} = \frac{1}{{\sqrt 2 }}{\left( {\begin{array}{*{20}{c}}
1&{\nu  i} \end{array}} \right)^t}$.
Then we have
$g_{\bf{q}}^{\nu ,{\nu ^\prime }} = \nu {\delta _{\nu ,\nu '}}$
and arrive at 
\begin{eqnarray}
{L^z} = \frac{1}{{2}}\sum\limits_{{\bf{q}},\nu } {\nu \varphi _{{\bf{q}}\nu }^\dag {\tau_2}\varphi _{{\bf{q}}\nu }^{}}.
\end{eqnarray}

Since we are interested in transfer of angular momenta between chiral phonons and localized spins, we consider the interaction between them in the form of spin-orbit coupling:
\begin{eqnarray}
V = \lambda {L^z}\sum\limits_{i = 1,2} \delta  \left( {{\bf{r}} - {{\bf{R}}_i}} \right)S_i^z
\end{eqnarray}
where the localized spins at ${\bf{R}}_1$ and ${\bf{R}}_2 $ are denoted by $S_1$ and $S_2$, respectively. 
The parameter $\lambda$ represents the coupling strength.
This interaction stems from Zeeman effect by the magnetic field due to chiral phonons. In 2D, local atomic rotation induces magnetic field perpendicular to the 2D plane. The magnetic field is proportional to phonon angular momentum. Therefore, the $z$-components of spins and phonon angular momentum are involved.

The spin-spin interaction can be calculated by second-order correction with respect to $V$ to the free energy as follows\cite{RKKY,Kogan,Oriekhov}.
The free energy of the system is given by \cite{Altland,Nagaosa}
\begin{equation}
F=T \operatorname{Tr} \ln \left[-D^{-1}+V\right]
\end{equation}
which can be expanded with respect to $V$ as
\begin{equation}
F \simeq  T \operatorname{Tr} \ln \left[-D^{-1}\right]  +\frac{1}{2}T \operatorname{Tr}\left[D V D V\right]+...
\end{equation}
Here, $\operatorname{Tr}$ means the integration or summation over modes, coordinates, frequences and $\tau$ space.
By inserting Eq.(7), we arrive at the second-order correction to the free energy as
\begin{widetext}
\begin{eqnarray}
\delta {F_{}} = \frac{{{\lambda ^2}}}{8}S_1^zS_2^zT\sum\limits_{n,\nu} {{\mathop{\rm tr}\nolimits} } \left[ {{\tau_2}D\left( {{{\bf{R}}_1} - {{\bf{R}}_2},{\omega _n},\nu} \right)} \right.\left. {{\tau_2}D\left( {{{\bf{R}}_2} - {{\bf{R}}_1},{\omega _n},\nu} \right)} \right].
\end{eqnarray}
Here, ${\rm tr}$ means the trace in  $\tau$ space.

We first consider acoustic phonons with $\omega _{\bf{q}\nu} = {v_a}q$ and $q = \left| {\bf{q}} \right|$. 
Then, the components of the phonon Green function $D$ are obtained as
\begin{eqnarray}
{D_{0}}({\bf{R}},\omega _n^{},\nu) = - \frac{1}{{2\pi {v_a}}}\left( {\frac{1}{R} - \frac{{\pi \left| {\omega _n^{}} \right|}}{{2{v_a}}}{I_0}\left( {\frac{{\omega _n^{}}}{{{v_a}}}R} \right) + \frac{{\pi  {\omega _n^{}} }}{{2{v_a}}}{L_0}\left( {\frac{{\omega _n^{}}}{{{v_a}}}R} \right)} \right),\\ {D_{2}}({\bf{R}},\omega _n^{},\nu)  = \frac{{\omega _n^{}}}{{2\pi v_a^2}}{K_0}\left( {\frac{{\left| {\omega _n^{}} \right|}}{{{v_a}}}R} \right)
\end{eqnarray}
with $R = \left| {\bf{R}} \right|$ and ${\bf{R}}={{\bf{R}}_1} - {{\bf{R}}_2}$. Here, $I_0$ and $K_0$ are the modified Bessel functions and $L_0$ is the Struve function.
\end{widetext}
Here, we have used the following relations:\cite{Prudnikov}
\begin{equation}
J_0(x)=\frac{1}{2 \pi} \int_0^{2 \pi} e^{i x \cos \varphi} d \varphi, \quad K_0(a b)=\int_0^{\infty} \frac{x J_0(a x)}{x^2+b^2} d x,
\end{equation}
\begin{equation}
\int_0^{\infty} \frac{x^2 J_0\left(c x\right)}{x^2+d^2} d x=\frac{1}{c}-\frac{\pi}{2}|d| I_0(cd)+\frac{\pi}{2} d L_0(cd)
\end{equation}
where $a, b,c>0$ and $J_0(x)$ is the Bessel function.

\begin{figure}[htb]
\begin{center}
\includegraphics[clip,width=8cm]{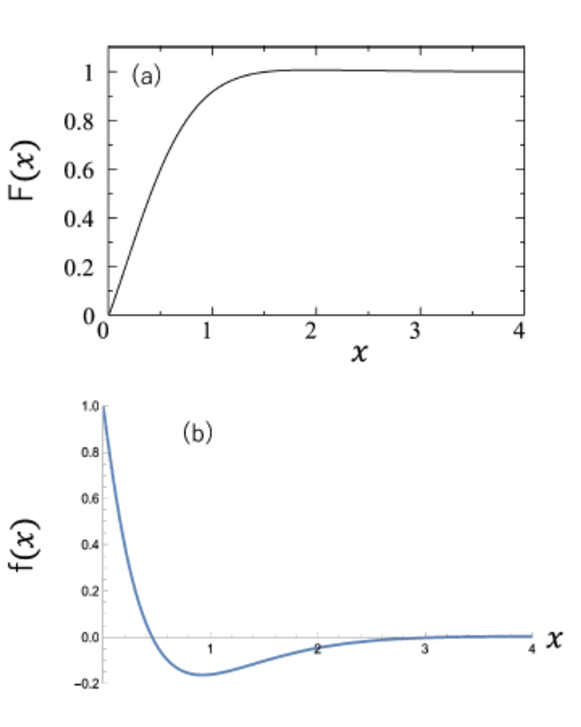}
\end{center}
\caption{
Plots of (a) $F(x)$ and (b)  $f(x)$.}
\label{f2}
\end{figure}

Therefore, we arrive at the expression of the correction to the free energy:
$\delta {F_{}} = J_z^aS_1^zS_2^z$ with
\begin{eqnarray}
J_z^a = \frac{{{\lambda ^2}}}{{8{\pi ^2}v_a{R^3}}}xF(x),\; F(x) = \sum\limits_n {f(2nx)}, \\ f(x) = {{\left( {1 - \frac{{\pi \left| x \right|}}{2}{I_0}\left( x \right) + \frac{{\pi x}}{2}{L_0}\left( x \right)} \right)}^2} - {{\left( {x{K_0}\left( x \right)} \right)}^2}
\end{eqnarray}
with $x = \frac{T}{{{v_a}}}R$.

Figure \ref{f2} (a) shows the behavior of $F(x)$. It increases linearly with $x$ for $x<1$ and saturates at $x \sim 1$. Therefore, for $x<1$, i.e., $R<\xi_T$ where $\xi_T=v_a/T$ is the thermal coherence length,  the exchange coupling $J_z^a$ behaves as 
\begin{eqnarray}
J_z^a \sim T^2/R
\end{eqnarray}
while for $x>1$ ($R>\xi_T$), it behaves as 
\begin{eqnarray}
J_z^a \sim T/R^2.
\end{eqnarray}
Since $F(x)>0$, we see that the exchange couplng is positive.

Figure \ref{f2} (b) shows the behavior of $f(x)$. It oscillates as a function of $x$ and goes to zero for $x \gg 1$. Thus, we find that the intergers $n$ which satisfy $2 n x \lesssim 3$ dominantly contribute to $J_z^a$ (see Eq.(15)).

Let us  estimate the magnitude of the exchange interaction for acoustic phonons. For $\lambda=10$meV/nm$^2$, $T=300$K, ${v_a}=10^4$m/s, and $R=1$nm, we have
$x \sim 4$ and $J_z^a \sim 10^{-2} {\rm{eV}}$.

\begin{figure}[htb]
\begin{center}
\includegraphics[clip,width=8cm]{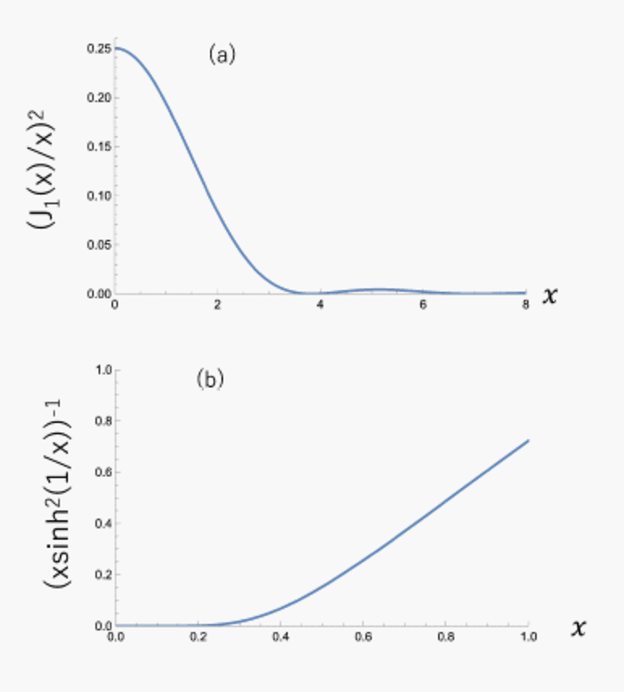}
\end{center}
\caption{
Plots of (a) $(J_1(x)/x)^2$ which shows the dependence of $J_z^o$ on $R$ and (b) $(x \sinh^{2}(1/x))^{-1}$ which exhibits the dependence of $J_z^o$ on $T$.  }
\label{f3}
\end{figure}

Next, let us consider optical phonons with $\omega _{\bf{q}\nu} = \omega _0$ (constant). 
Then, the components of the phonon Green function $D$ are obtained as\cite{Prudnikov} 
\begin{eqnarray}
{D_{0}}({\bf{R}},\omega _n^{},\nu)  = - \frac{1}{{2\pi }}\frac{{\omega _0^{}}}{{\omega _n^2 + \omega _0^2}}\frac{{\Lambda {J_1}\left( {\Lambda R} \right)}}{R},\\ {D_{2}}({\bf{R}},\omega _n^{},\nu)  = \frac{1}{{2\pi }}\frac{{\omega _n^{}}}{{\omega _n^2 + \omega _0^2}}\frac{{\Lambda {J_1}\left( {\Lambda R} \right)}}{R}
\end{eqnarray} 
where $\Lambda$ is the cutoff for $q$ and $J_1$ is the Bessel function. 

Therefore, we obtain the expression of the correction to the free energy:
$\delta {F_{}} = J_z^o S_1^zS_2^z$ with
\begin{eqnarray}
J_z^o = \frac{{{\lambda ^2}}}{{32{\pi ^2}}}\frac{{{{\left( {\Lambda {J_1}\left( {\Lambda R} \right)} \right)}^2}}}{{{R^2}T{{\sinh }^2}\left( {\frac{{\omega _0^{}}}{{2T}}} \right)}}.
\end{eqnarray}
We find that $J_z^o $ is positive.
From this expression, we find asymptotic behaviors of the exchange interaction at $T \to 0$ and $T \to \infty $:
\begin{eqnarray}
J_z^o \sim \frac{{{e^{ - \frac{{\omega _0^{}}}{{T}}}}}}{T}\; (T \to 0),\;J_z^o \sim T\; (T \to \infty )
\end{eqnarray}
and also those at $R \to 0$ and $R \to \infty $:
\begin{eqnarray}
J_z^o \sim {\rm{const.}}\; (R \to 0),\\ J_z^o \sim \frac{1}{{{R^3}}}{\cos ^2}\left( {\Lambda R - \frac{3}{4}\pi } \right)\; (R \to \infty ).
\end{eqnarray}
This oscillating behavior for sufficiently large $R$ has a similarity to the RKKY interaction. \cite{RKKY}

To see the dependence of $J_z^o$ on $R$, we show the behavior of $(J_1(x)/x)^2$ in Fig. \ref{f3} (a) (see also Eq.(21)). It is an oscillatory decreasing junction of $x$. 
To see the dependence of $J_z^o$ on $T$, we show the behavior of $(x \sinh^{2}(1/x))^{-1}$ in Fig. \ref{f3} (b). It is a monotonically increasing function of $x$.
The asymptotic behaviors in these figures are consistent with Eqs.(17-19).

Let us  estimate the magnitude of the exchange interaction for optical phonons. For $\lambda=10$meV/nm$^2$, $T=300$K, $\omega _0=10$meV, $\Lambda=10^{10}$/m, and  $R=1$nm,  we have
$J_z^o  \sim 10^{-4}$ eV.

Here, we provide several discussions on the results obtained in this paper:

(i) For both acoustic and optical phonons, the exchange interactions become zero at zero temperature and proportional to the temperature at high temperature. This is because these phonons are thermally excited and reflects the behaviors of the Bose distribution functions at these temperatures. 

(ii) We have assumed that the coupling strength between chiral phonons and spins $\lambda$ in Eq.(7) is a constant. In general, it depends on temperature.\cite{Gao2023} In this case, the temperature dependences of the exchange couplings are superimposed by that of $\lambda^2$.

(iii)
We have found that the exchange interactions mediated by chiral phonons show algebraic decay with respect to the distance between the two spins.
On the other hand, the exchange interaction mediated by electrons in insulators  shows exponential decay with the distance.\cite{Bruno,Faure}
Therefore, it is possible to distinguish between the contribution from chiral phonons and that from electrons by comparing the dependence of the exchange coupling on the distance between the localized spins. 

(iv)
Let us compare the exchange interactions mediated by chiral phonons and electrons in metals. As for the RKKY interaction, the exchange couplings are proportional to trigonometric functions of $k_F R$ where $k_F$ is the Fermi wavenumber. \cite{RKKY,Kogan,Oriekhov} These trigonometric functions lead to oscillations of the exchange couplings accompanied by their sign change with respect to $R$. 
In contrast, since phonons are bosons, the chemical potential is zero (correponding to $k_F=0$ in the case of electrons), and therefore we do not see such oscillations of the exchange couplings mediated by chiral phonons as we have seen in this paper.
The absence of  oscillations accompanied by sign change  is expected to be a general feature of exchange couplings mediated by bosons.

In summary, 
we have investigated interaction between two magnetic impurities on top of a two dimensional insulator in the presence of chiral phonons by second-order perturbation theory.
We have shown that this interaction arises from angular momentum of phonons through spin-chiral phonon interaction of the form of spin-orbit coupling.
Analytical expressions of the interaction for acoustic and optical phonons are obtained. The exchange interactions are always positive in contrast to the RKKY interactions due to the bosonic nature of phonons.
We have found that the exchange interactions for acoustic and optical phonons show power-law decay with respect to the distance between the two magnetic impurities and are proportional to the temperature of the system at high temperature.

This work was supported by JSPS KAKENHI Grant No.~JP30578216.

\end{document}